\documentclass{article}
\usepackage{smc}
\usepackage{times}
\usepackage{ifpdf}
\usepackage[english]{babel}
\usepackage{cite}
\usepackage{url}

\def\papertitle{An interactive music infilling interface for pop music composition}
\def\firstauthor{Rui Guo}

\newif\ifpdf
\ifx\pdfoutput\relax
\else
   \ifcase\pdfoutput
      \pdffalse
   \else
      \pdftrue
\fi

\ifpdf 
  \usepackage[pdftex,
    pdftitle={\papertitle},
    pdfauthor={\firstauthor},
    bookmarksnumbered, 
    pdfstartview=XYZ 
   ]{hyperref}

  \usepackage[pdftex]{graphicx}
  \graphicspath{{./figures/}}
  \DeclareGraphicsExtensions{.pdf,.jpeg,.png}

  \usepackage[figure,table]{hypcap}

\else 
  \usepackage[dvips,
    bookmarksnumbered, 
    pdfstartview=XYZ 
  ]{hyperref}  

  \usepackage[dvips]{epsfig,graphicx}
  \graphicspath{{./figures/}}
  \DeclareGraphicsExtensions{.eps}

  \usepackage[figure,table]{hypcap}
\fi

\hypersetup{
    colorlinks,%
    citecolor=black,%
    filecolor=black,%
    linkcolor=black,%
    urlcolor=black
}

\title{\papertitle}

\oneauthor
  {\firstauthor}{University of Sussex \\ %
    {\tt \href{mailto:R.Guo@sussex.ac.uk}{R.Guo@sussex.ac.uk}}}

\begin{document}
\capstartfalse
\maketitle
\capstarttrue
\begin{abstract}
Artificial intelligence (AI) has been widely applied to music generation topics such as continuation, melody/harmony generation, genre transfer and music infilling application. Although with the burst interest to apply AI to music, there are still few interfaces for the musicians to take advantage of the latest progress of the AI technology. This makes those tools less valuable in practice and harder to find its advantage/drawbacks without utilizing them in the real scenario. This work builds a max patch for interactive music infilling application with different levels of control, including track density/polyphony/occupation rate and bar tonal tension control. The user can select the melody/bass/harmony track as the infilling content up to 16 bars. The infilling algorithm is based on the author's previous work, and the interface sends/receives messages to the AI system hosted in the cloud. This interface lowers the barrier of AI technology and can generate different variations of the selected content. Those results can give several alternatives to the musicians' composition, and the interactive process realizes the value of the AI infilling system.
\end{abstract}
%


\section{Introduction}

Most of the work on AI music generation provides a google colab notebook for exploration\cite{huang2018music}. That notebook hides the technical details and makes it easier to control the system, however, it is still far from an intuitive user experience. The survey \cite{Huang2020AISC} shows the musicians prefer more control rather than generating end-end music.  The survey suggests there are several gaps between musicians and those algorithms, in the concept of ``disposable'', ``context-aware'' and ``steerable''. 

To fill in that gap, this work builds a Max patch integrated with an AI pop music infilling model with multiple level control parameters. Given a MIDI file with one to three tracks in the types of melody, bass and harmony, the user can select a whole track/several bars and have new variations given the surrounding music, which is to have an ``infilling'' of the selected music region. The track/bar control parameters can help the user to steer the music to explore different textures, and tonal tension control can change the total tension curve by drawing a line. The user can compose a pop song by interactively repeating the generation process in this interface with ease.


\section{Motivation}

This work is based on a pop music infilling system with multiple levels of control\cite{guo2022musiac}. That system can steer each track's note density/polyphony/occupation level and bar tonal tension\cite{herremansTensionRibbonsQuantifying}. The original interface in \cite{guo2022musiac} is a colab notebook, and to make it more accessible, a Max patch is prepared to communicate with the deep learning infilling system and better fit the composer's writing process.  Compared with other interfaces\cite{hadjeres2021piano}, this one provides both track and bar control. It is also disposable and context-aware by the nature of music infilling. The aim of this interface is to be compatible with any music DAW that can receive MIDI messages, and easy to set the infilling region and control parameters. It should also play/display the music notation in the interface. The target user of this interface is the music composers of all levels. 

\section{Interface}

\begin{figure}[t]
\centering
\includegraphics[width=0.6\columnwidth]{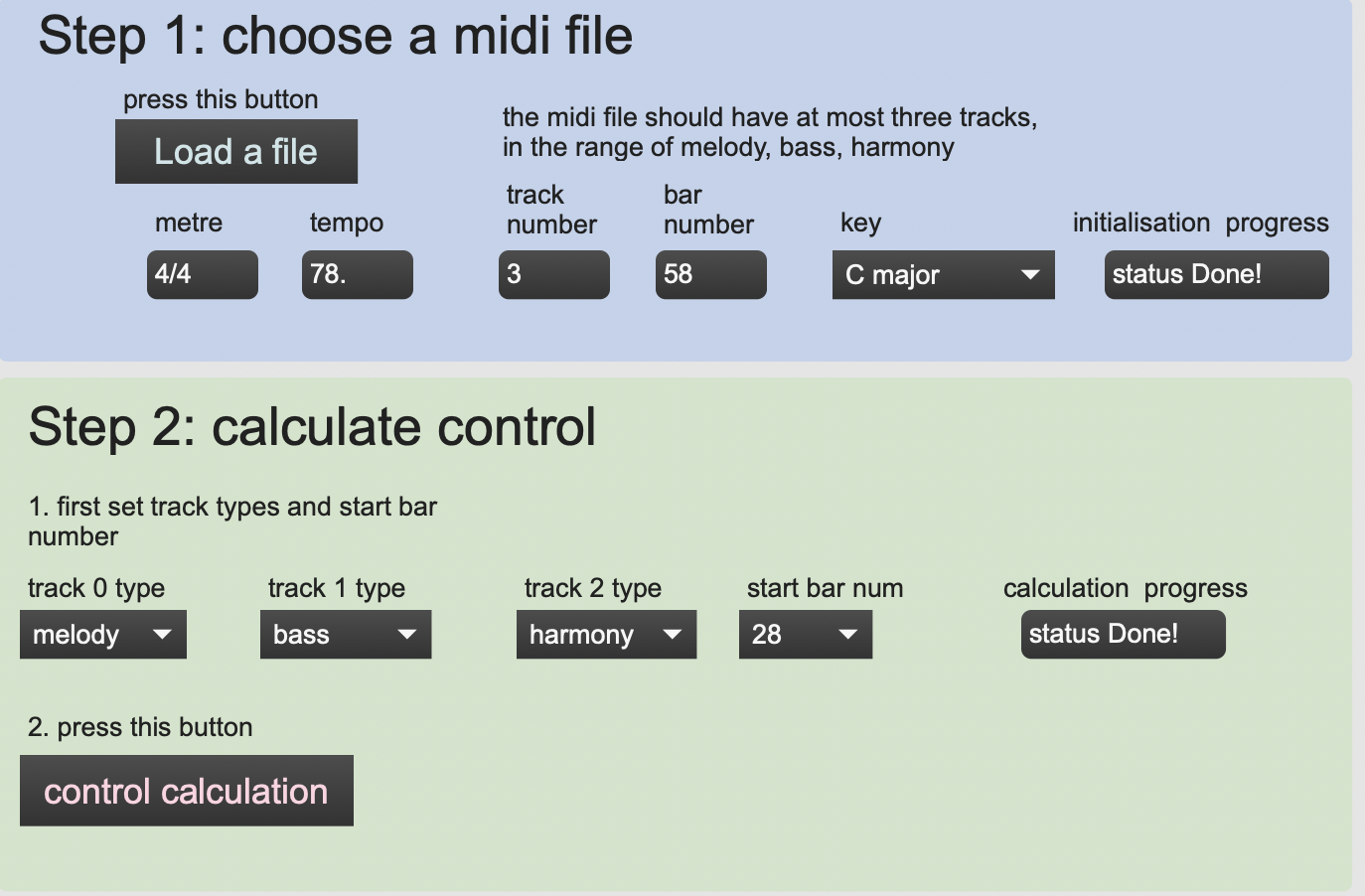}
\caption{Input MIDI selection and track type settings\label{fig:selection}}
\end{figure}

The input of this interface is MIDI files and it will use the first three tracks as input to the system. The tracks should be in the types of melody, bass and harmony with arbitrary order, but each type can only be in one track. After clicking the ``Load a file'' button in \figurename~\ref{fig:selection}, the MIDI information including track number, bar number, metre, tempo and key are displayed. The key calculation is based on the tonal tension theory and also the music21\cite{music21} key estimation. 

\begin{figure}[t]
\centering
\includegraphics[width=0.6\columnwidth]{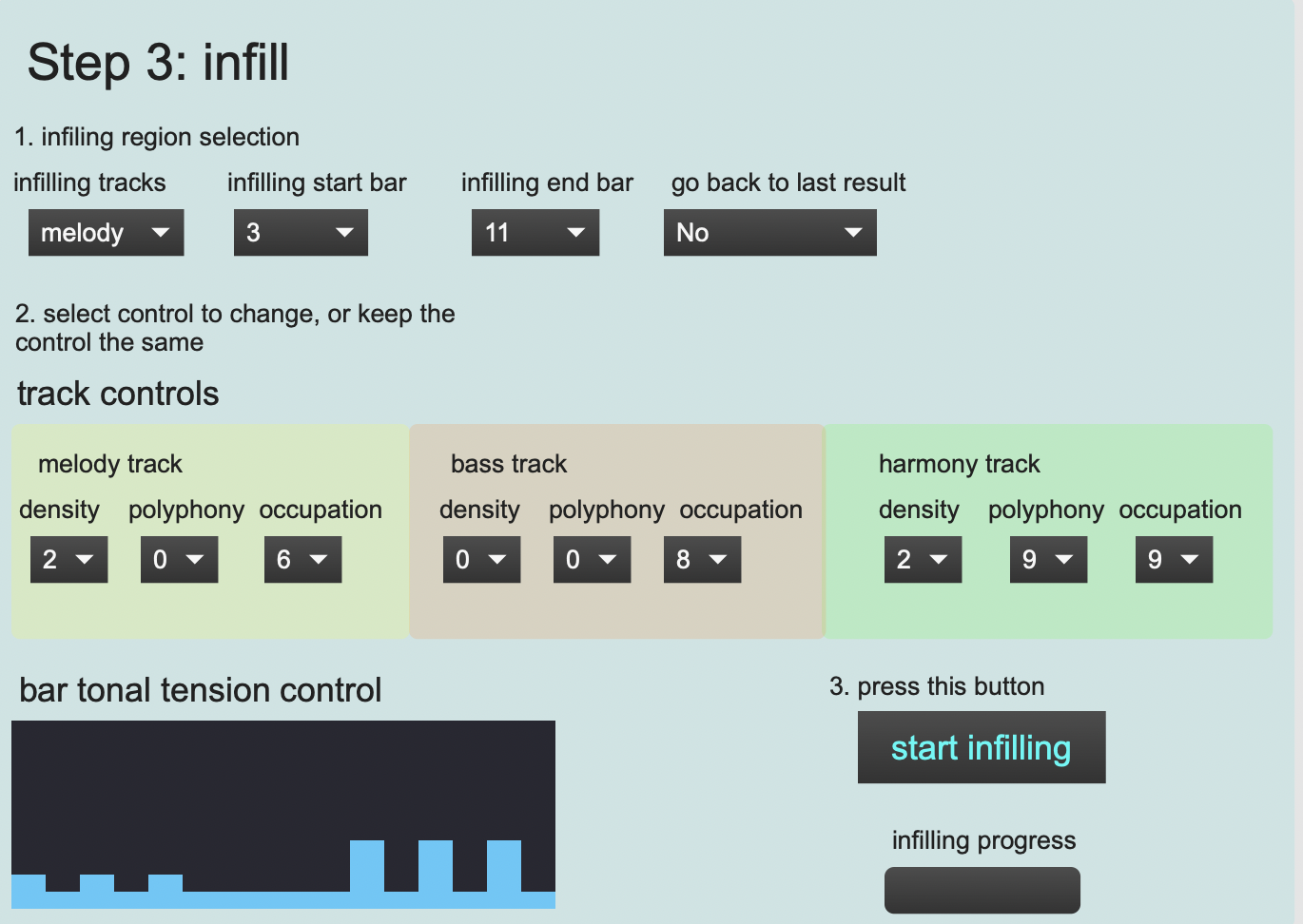}
\caption{Music infilling settings. Arbitrary sections in the music can be selected for infill, and the track/bar control parameters can be adjusted.\label{fig:infill}}
\end{figure}

Before the infilling process, each track's type should be set to one of the ``melody'', ``bass'', ``harmony'' or ``empty''. Because the input music length can be arbitrary, the user needs to set a start bar of music if the total bar length is more than 16, which is the bar limit of the proposed infilling system. To process the whole song, the user can apply this tool to any 16 bar sections of the whole song. After clicking the ``Calculating control'' button, the track/bar control parameters of the region from the selected start bar are calculated and displayed in the \figurename~\ref{fig:infill}. 

The infilling process will assume the music has three tracks. If the input music has less than three tracks, the empty track can also be infilled in the infilling stage. The minimum infilling region of this application is one track in a bar, e.g. melody track in bar 5. A whole bar can be infilled by setting ``infilling tracks'' to ``all''.  Each track has its own control, and each bar tonal tension control can also be altered.  The track parameters are in the range of level 0-9, and the bar tonal tension is set by dragging the slider according to that bar. The user can also drag a tension curve and infill those selected bars.

\begin{figure}[t]
\centering
\includegraphics[width=0.6\columnwidth]{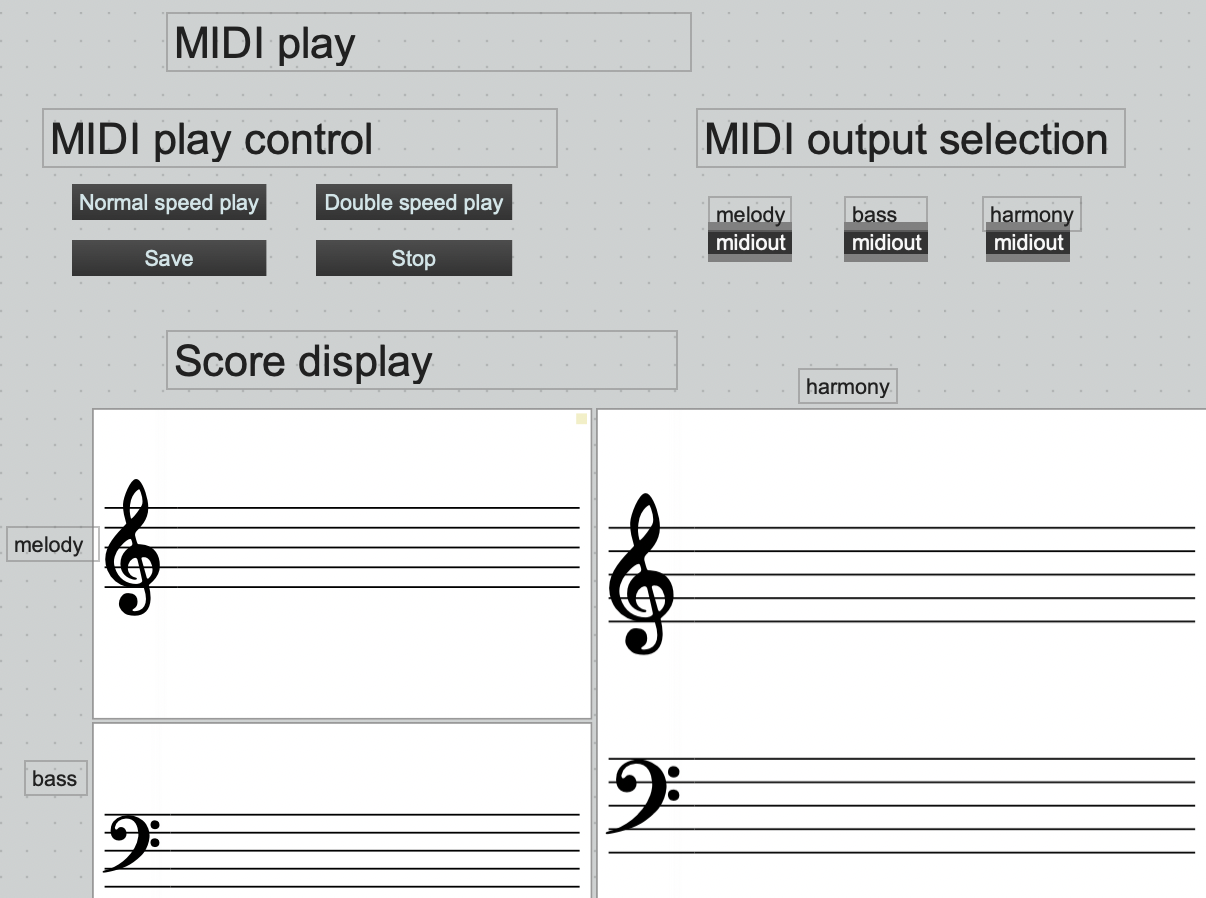}
\caption{The play control and music notation display. The result MIDI can be sent to different DAWs for further edit or saved. Each track is in a separate music sheet.\label{fig:sheet}}
\end{figure}

After clicking the ``start infilling'' button, the return result can be played/saved by the play control in \figurename~\ref{fig:sheet}. The bach\cite{agostini2015max} library is employed for music notation display and each track is in a separate music sheet. The user can send the tracks to different DAWs by double clicking the ``midiout'' button.

The node.script module is utilised in Max to communicate with a PyTorch model served in the Flask framework. Only the MIDI file and the control information is sent/received with little bandwidth requirement. The model is served in cloud with GPU and can be also installed on private servers if needed. The Max patch is shared for free exploration\footnote{\url{https://github.com/ruiguo-bio/max_infilling}} 

\section{Infilling examples}

A starting point can be a complete MIDI file with three tracks and at least 16 bars. The infilling function is applied iteratively to change part of the music. From the author's experience, the infilling result is better if working on a smaller scale with more surrounding context, e.g. to infill 8 bars of melody or 4 whole bars at the beginning of music. If set the ``go back to last result'' to ``No'', the next infilling operation will be based on this result, i.e. keep the latest result as the starting point.

A MIDI with less than three tracks and only a few bars can also be used to build an entire song. If the input music has only a melody track, by selecting the control parameters of the bass/harmony tracks, those tracks can also be infilled according to the control. Due to the sampling operation in the generation, the generated result generally follows the control, but it is not ensured the result control level is always the same as the settings.

\section{Conclusions}
This work builds a Max patch to integrate an AI music infilling model with an intuitive interface. The infilling region can range from a track in a bar to the whole song. The track/bar level control helps to steer the generation process, and the generated MIDI can be sent to different DAWs for further edit. This tool can give numerous variations of the input and inspire the composer in the music writing.


\bibliography{smc2022bib}

\end{document}